\documentclass[aps,twocolumn,showpacs,preprintnumbers,prl,superscriptaddress]{revtex4}

\usepackage[dvips]{graphicx}
\usepackage{amssymb, amsmath}

\begin{document}

\title{Accuracy of basis-set extrapolation schemes for DFT-RPA correlation energies in molecular calculations}
\author{Eduardo Fabiano}
\affiliation{National Nanotechnology Laboratory (NNL), Istituto Nanoscienze-CNR,
Via per Arnesano 16, I-73100 Lecce, Italy }
\author{Fabio Della Sala}
\affiliation{National Nanotechnology Laboratory (NNL), Istituto Nanoscienze-CNR,
Via per Arnesano 16, I-73100 Lecce, Italy }
\affiliation{Center for Biomolecular Nanotechnologies @UNILE, Istituto Italiano di Tecnologia (IIT), Via Barsanti, 73010 Arnesano (LE), Italy}

\begin{abstract}
We construct a reference benchmark set for atomic and molecular 
random-phase-approximation (RPA) correlation energies in a density
functional theory (DFT) framework at the complete basis 
set limit. This set is used to evaluate the accuracy of some 
popular extrapolation schemes for RPA 
all-electron molecular calculations. The results indicate that
for absolute energies accurate results, clearly
outperforming raw data, are achievable with two-point extrapolation 
schemes based on quintuple- and sextuple-zeta
basis sets. Moreover, we show that results in good agreement with the
benchmark can also be also obtained by using a semiempirical extrapolation 
procedure based on quadruple- and quintuple-zeta basis sets.
Finally, we analyze the performance of different extrapolation
schemes for atomization energies.
\end{abstract}

\maketitle

\section{Introduction}
\label{sec_int}
The random phase approximation (RPA) for electron correlation was originally
developed in the context of the many-body perturbation treatment
of the uniform electron gas \cite{bohm53,gellmann57} and later 
reformulated in the framework of density functional theory (DFT) via
the adiabatic-connection fluctuation-dissipation theorem 
\cite{langreth77}. 
In the last years, 
RPA correlation witnessed an increasing interest in quantum chemistry
\cite{furche01,fuchs02,furche05,dobson05,scuseria08,gruneis09,janesko09_ref,janesko09,toulouse09,toulouse10,ruzsinszky10,nguyen10,paier10,eshuis10,jansen10,ren11,eshuis11,ruzsinszky11,angyan11,hess11t,eshuis12,paier12,ren12,ren12_2},
due to its ability to describe with 
good accuracy a rather large variety of reaction and interaction 
energies.
In this context RPA is generally employed as a post
Kohn-Sham (KS) approach where the exact exchange (EXX) and 
the correlation energies
are evaluated using the eigenvalues and eigenfunctions from a KS calculation 
with conventional exchange-correlation (XC) functionals.
EXX KS orbitals \cite{hess11prl,hess10,hess11m} have been 
also employed via the optimized effective potential (OEP) method,  while  for  
a fully variational approach self-consistent EXX-RPA orbitals 
are required 
\cite{godby86,kotani,niquet03,grun06,hell07,hell10,hell12,verma12}. 
In all cases practical applications of the method are often hindered by the
the very slow convergence of the RPA-correlation energy with the basis set
dimension \cite{furche01,harl08,eshuis12_2}, 
which is related to the electron cusp problem
\cite{furche01}, in analogy with wavefunction methods. 
The cusp problem is strongly reduced for range-separated RPA 
\cite{janesko09,toulouse09,toulouse10,janesko09_3,zho10}, whereas 
for full-range RPA approaches 
extrapolation to the complete basis set (CBS) must be employed. 

Basis set extrapolation to the CBS limit has been longley studied for
post Hartree-Fock correlated methods, with the aim of reducing as
much as possible the basis set truncation error without resorting to
the brute force employment of extremely large basis sets.
Numerical work 
\cite{schwartz62,schwartz63,carroll79,schmidt83,hill85,kutzelnigg92,schmidt94} 
on various correlation treatments of the helium atom
has shown that the basis set incompleteness error is approximately
proportional to the inverse third power of the maximum angular momentum,
provided that all angular momenta are radially saturated.
However, the ideal conditions of fully saturated angular momenta in an
atomic system are rather distant from those encountered in real applications
on molecules with finite basis sets. Nevertheless, the correlation energy
has been empirically observed to converge regularly enough to allow
the development of various accurate extrapolation schemes which
have been assessed in several studies in literature
\cite{muller,feller11,bakowies07,bakowies07_2,varandas07}. 
On the other hand, no such study has been performed for RPA correlation 
energies. 

In this paper we aim at filling this gap and shed light on the 
usefulness and limits of different extrapolation schemes for 
RPA correlation energies.
In this respect the goal of the present work is to: (1) create
a set of benchmark CBS RPA absolute correlation energies to
be used as reference for future assessments; (2) validate some
extrapolation schemes, usually applied in the context of coupled cluster
theory, for the calculation of CBS RPA absolute energies, including 
semiempirical techniques to approach the extrapolation from relatively
small basis sets; (3) test the convergence behavior of 
different basis sets and the reliability of the extrapolations 
schemes for atomization energies, which exploit a significant error 
cancellation effect. Absolute correlation energies will be considered,
as in many corresponding studies of correlated methods, because
these converge regularly with the basis set and provide an ideal 
quantity for the study of extrapolation schemes.
In this paper we focus on light elements because they provide a simpler
convergence of the correlation energy and can be more easily
investigated with large basis sets.

\section{Computational details}
\label{sec_compdet}
RPA calculations were performed using the eigenvalues and the
orbitals from DFT calculations based on the Perdew-Burke-Ernzerhof
(PBE) exchange-correlation functional \cite{pbe}.
In all calculations Dunning's correlation-consistent 
basis sets \cite{dunning89,woon94,feller99,feller00} 
(cc-pV$n$Z, $n=D,5,6,7$) augmented with
core and core-valence basis functions \cite{woon95,peterson02,hill10}
were employed. These are hereafter denoted as V$n$Zcv with
$n$=4, 5, 6, and 7.
Augmented diffuse basis functions were not considered in the present work
since they were shown to slow down RPA basis set convergence without
bringing substantial benefits in most cases \cite{eshuis12_2}.
Also, no basis set superposition error correction was considered for
atomization energies.
For molecules experimental geometries were considered 
\cite{boyd55,venkateswarlu55,redington62,herzberg66,sverdlov74,tsuboi75,hirota79,hoy79,huber79,herbst84,junttila94,gurvich89,kuchitsu98,nist}.
All calculations were performed with the TURBOMOLE
program package \cite{turbomole} using the implementation described in Ref. \cite{furche01}.

\begin{table*}
\begin{center}
\caption{\label{tab1}RPA correlation energies of several atoms and molecules
for different basis sets and extrapolation schemes. The last two columns report the
best estimate (average of the extrapolated values) and the estimated uncertainty $\delta$. 
At the bottom of the table we report for the raw data from different basis sets the
mean error (ME), mean absolute error (MAE), mean absolute relative error (MARE),
maximum absolute deviation (MAD), and standard deviation with respect to the best estimate.
All results are in mHa.}
\begin{ruledtabular}
\begin{tabular}{lrrrrrrrrr}
System & V5Zcv & V6Zcv & V7Zcv & V7Zcv+ & $1/(n+d)^3$ & $1/(n+d)^4$ & $1/n^\alpha$ & Best est. & $\delta$ \\
\hline
H &  -20.3 &   -20.6 &   -20.7 & -20.9  &  -21.0 &   -21.0 &   -21.0 &   -21.0 & 0.1 \\
C & -284.9 &  -288.5 &  -290.1 & -292.0 & -292.6 &  -293.0 &  -292.9 &  -292.8 & 0.7 \\
N & -327.8 &  -332.8 &  -334.9 & -336.8 & -337.2 &  -337.5 &  -337.1 &  -337.3 & 0.6 \\
O & -416.5 &  -423.4 &  -426.7 & -430.4 & -431.5 &  -432.0 &  -431.9 &  -431.8 & 0.9 \\
F & -504.1 &  -513.1 &  -517.5 & -522.0 & -523.9 &  -524.6 &  -524.4 &  -524.3 & 1.3 \\
Ne & -579.2 &  -590.5 &  -595.6 & -600.5 & -601.9 &  -602.6 &  -602.0 &  -602.2 & 1.3 \\
H$_2$ & -80.0 & -80.5 &	-80.7 &          & -81.2 & -81.1 & -81.2 & -81.2 & 0.2 \\
NH & -399.3 &  -404.8 &  -407.2 &       & -410.4 &  -410.7 &  -410.4 &  -410.5  & 0.6 \\
NH$_2$ & -468.1 &  -474.1 &  -476.8 &        & -480.4 &  -480.9 &  -480.6 &  -480.6 & 0.7 \\
CH$_4$ & -491.4 &  -496.3 &  -498.6 &       &  -501.4 &  -501.7 &  -501.4 &  -501.5 & 0.6 \\
NH$_3$ & -533.4 &  -539.3 &  -541.8 &       &  -544.5 &  -544.9 &  -544.5 &  -544.6 & 0.7 \\
H$_2$O & -565.7 &  -573.8 &  -577.4 &        & -581.7 &  -582.3 &  -581.8 &  -581.9 & 0.9 \\
FH & -582.1 &  -591.9 &  -596.3 &        & -602.0 &  -602.7 &  -602.2 &  -602.3 & 1.1 \\
C$_2$H$_2$ & -756.5 &  -765.3 &  -769.4 &       &  -774.7 &  -775.3 &  -774.9 &  -775.0 & 1.1 \\
CN & -774.5 &  -783.9 &  -787.8 &        & -792.3 &  -792.9 &  -792.3 &  -792.5 & 1.0 \\
HCN & -799.8 &  -809.6 &  -813.8 &       &  -818.7 &  -819.3 &  -818.7 &  -818.9 & 1.0 \\
CO & -821.6 &  -832.8 &  -837.5 &        & -842.8 &  -843.5 &  -842.7 &  -843.0 & 1.2 \\
N2 & -833.6 &  -844.6 &  -849.3 &        & -855.1 &  -855.8 &  -855.2 &  -855.4 & 1.2 \\
C$_2$H$_4$ & -833.4 &  -842.5 &  -846.7 &       &  -852.3 &  -852.9 &  -852.5 &  -852.6 & 1.1 \\
HCO & -865.9 &  -877.5 &  -882.5 &        & -888.4 &  -889.1 &  -888.4 &  -888.6 & 1.2 \\
H$_2$CO & -909.8 &  -921.6 &  -926.7 &       &  -932.7 &  -933.5 &  -932.7 &  -933.0 & 1.2 \\
O$_2$ & -973.1 &  -987.6 &  -993.9 &        & -1001.5 &  -1002.4 &  -1001.6 &  -1001.8 & 1.5 \\
H$_3$COH & -984.0 &  -996.6 &  -1002.1 &       &  -1008.7 &  -1009.5 &  -1008.8 &  -1009.0 & 1.3 \\
HOOH & -1077.2 &  -1092.6 &  -1099.4 &       &  -1107.6 &  -1108.6 &  -1107.7 &  -1108.0 & 1.6 \\
F$_2$ & -1125.2 &  -1143.9 &  -1152.3 &        & -1163.0 &  -1164.3 &  -1163.4 & -1163.6 & 2.4 \\
        &      &      &      &    &         &        &          & & \\
ME      & 17.9 & 9.0 & 5.1   &    &           0.2   &   -0.4   & 0.1 & & \\
MAE	& 17.9 & 9.0 & 5.1   &    &           0.2   &   0.4    & 0.1 & & \\
MARE	&2.72\%&1.40\%&	0.80\%&   &          0.04\% &   0.05\% & 0.02\% & & \\
MAD	& 38.4 & 19.7 &	11.3 &    &           0.5   &   0.7    & 0.3 & & \\
Std. Dev.& 8.7 & 4.4 & 2.5  &     &           0.1   &   0.2    & 0.1 & & \\
\end{tabular}
\end{ruledtabular}
\end{center}
\end{table*}

\section{Reference data}
The construction of a set of benchmark CBS-limit RPA energies is a fundamental
step before any possible assessment work on extrapolation schemes.
Unfortunately, in the case of RPA correlation energies, this is a difficult 
task. In fact, unlike for MP2 or coupled cluster methods, no explicitly 
correlated \cite{kutzelnigg85} RPA data exists that can serve as an accurate
reference. At the same time, a brute force strategy, based on the use
of very large basis sets, is hampered by the extremely slow
convergence rate of the RPA calculations. Thus, in order to
approach the true CBS limit as close as possible and provide accurate 
reference CBS energies, some form of extrapolation is necessary.
However, this introduces inevitably an undesirable degree of uncertainty
in the data that can be anyway reasonably reduced by an appropriate
control of the possible sources of errors.

In this work we construct our set of reference RPA correlation energies
based on two leading criteria: (1) we employ as a basis for the 
extrapolation the best results at our disposal (i.e. those from 
V5Zcv to V7Zcv calculations),
in order to recover as much as possible the ``theoretical'' asymptotic
converge of the correlation energy that underlies all extrapolation
formulas and thus reduce the extrapolation error. 
For this reason we prefer to exclude data from V4Zcv 
calculations (or lower) that may increase the computational noise due to the 
incompleteness of the basis set. (2) We fix our best estimate of
the RPA correlation energy by averaging over the data obtained from several 
flexible three-parameters extrapolation formulas, in order to avoid
the possible bias characteristic of a specific extrapolation scheme.
A similar procedure was recently applied for the construction
of a benchmark set of CCSD(T) CBS energies \cite{feller11}.
In addition, we consider in detail the evolution of different 
extrapolated energies with dimension of the basis sets to fix
reasonable bounds to our uncertainty.

In more detail, we consider the popular extrapolation formula
\cite{feller11,helgaker97,halkier98}
\begin{equation}\label{e1}
E_n = E_\infty + \frac{A}{(n+d)^3} 
\end{equation}
and its variant proposed by Martin \cite{martin96}
\begin{equation}\label{e2}
E_n = E_\infty + \frac{A}{(n+d)^4} \ ,
\end{equation}
where $A$ and $d$ are parameters and $E_\infty$ is the CBS limit of the 
correlation energy. The free parameter $d$ accounts for the incompleteness
of the basis set angular-momentum saturation \cite{bakowies07} and
also partially introduces effectively higher-order contributions to the
asymptotic correlation expansion, as is readily recognized by
considering the alternative form for Eq. (\ref{e1}) 
\begin{equation}\label{e3}
E_n = E_\infty + \frac{A}{(n)^3} + \sum_{i=1}^\infty(-1)^i\binom{n}{i}\frac{Ad^i}{n^{3+i}}\ .
\end{equation}
In addition, we use the extrapolation formula proposed by Bakowies
\cite{bakowies07,bakowies07_2}
\begin{equation}\label{e4}
E_n = E_\infty + \frac{A}{n^\alpha}\ ,
\end{equation}
with $A$ and $\alpha$ adjustable parameters. Again, $\alpha$ provides 
an effective correction for deviations from the leading
asymptotic behavior $\propto n^{-3}$ \cite{bakowies07},
by partially adding high-order terms to the expansion.

The RPA correlation energies of several atoms and molecules as
resulting from V5Zcv, V6Zcv, V7Zcv calculations and 
the extrapolations with formulas (\ref{e1})-(\ref{e4}) 
are reported in Tab. \ref{tab1},
together with our best estimate for the CBS limit
RPA correlation energy.
All the extrapolation methods yield very close results,
with differences $<$ 1 mHa, justifying our assumption of 
taking as best estimate their average value. 

\begin{figure}
\includegraphics[width=0.9\columnwidth]{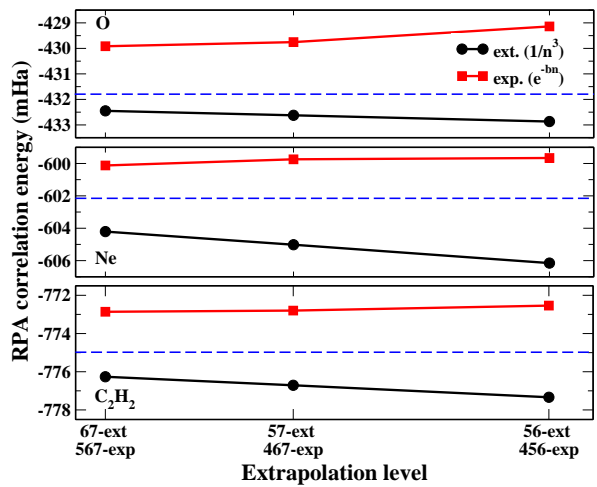}
\caption{\label{fig1} Extrapolated RPA correlation energies (mHa) of the O and Ne atoms and the C$_2$H$_2$ molecule as a function of different combinations of basis sets. The label $nm$-ext denotes an extrapolation based on Eq. (\ref{e1}) with $d=0$ and using V$n$Zcv and V$m$Zcv basis sets; the label $nmk$-exp denotes an exponential extrapolation using V$n$Zcv, V$m$Zcv, and V$k$Zcv basis sets. The dashed lines indicate our best estimate for the CBS RPA energy. The $x$-axis uses an inverse cubic scale.}
\end{figure}

We define in addition, in the last column of Tab \ref{tab1}, 
the error $\delta$ associated with of our benchmark  
energies in the following way. First, we note
that numerical evidence shows
that for the two limiting cases of an inverse cubic 
extrapolation (i.e. using Eq. (\ref{e1}) with $d=0$) and
of an exponential extrapolation, the exact CBS RPA correlation 
energy is approached from below and above, respectively. 
This is shown for some representative cases in Fig. \ref{fig1}, 
where the results of the two extrapolations with increasingly
large basis sets are reported,
and can be rationalized as follows: on one hand the
exponential extrapolation is well known to underestimate the CBS limit
\cite{muller}
because of the too fast rate of variation with the basis set
(exponential rather than with a power law), on the other hand
the simple inverse cubic extrapolation is likely to overestimate
the CBS limit because the 1/$n^3$ behavior is strictly valid only
asymptotically, while for real operative conditions a faster rate
of variation with the basis set shall be expected (hence the
$d$ parameter in Eq. (\ref{e1})).
The best extrapolated values obtained with this two procedures can 
thus be considered as upper and lower bounds for the true CBS RPA 
correlation energy. Then, after an analysis 
of the trends of Fig. \ref{fig1}, we assume as an 
estimation of the error half of the maximum difference between 
our best estimate and any of the two bounds for the energy. 

The reference energies of Tab. \ref{tab1} have thus an estimated accuracy of
about 1 mHa or less, except for the last four entries of the table,
which display slightly larger errors, with a maximum of 2.4 for F$_2$. 
We note however, that these systems also have the largest absolute 
correlation energies. In fact, the relative errors for all systems are very low
ranging between 0.11\% and 0.25\%, with a mean absolute relative error 
of 0.16\%.

To further validate our results we considered for the atoms also RPA
calculations with an extremely large basis set (V7Zcv+), constructed
by adding to the V7Zcv basis set centered on the atom additional 
uncontracted basis functions (up to angular momentum $g$) centered at 
six points located on the Cartesian axes at $\pm$1 Bohr of 
distance from the atom, for a total of 1113 basis functions per atom
(723 for the H atom). The resulting basis set allows to describe
effectively very high angular momentum contributions, 
thanks to the composition of momenta from functions at different centers,
without the need to effectively include basis functions with 
such high values of angular momentum into our basis set and partially avoiding
problems related to the linear dependence of basis functions. 
The results of the V7Zcv+ calculations, reported in Tab. \ref{tab1},
show that indeed by using a very large basis set the estimated
RPA CBS energy are approached  well, with deviations that are always
close  or lower than the expected uncertainty of the reference values. 
This result provides a good support  
for the accuracy of our estimation of the CBS limit of 
the RPA correlation as well as for the the corresponding errors.

The procedure outlined in this section, and whose results are reported in Tab.
\ref{tab1}, allows therefore to produce an accurate set of CBS RPA
correlation energies, improving of several mHa (about 5 to 10) with respect to
V7Zcv results (see bottom part of Tab. \ref{tab1}). 
Even more important improvements, up to 20 mHa, are found
with respect to V6Zcv calculations. The proposed set, despite
its limited dimension, provides in addition a representative selection
of molecules constituted from first-row atoms, including doublet and
triplet ground states. Thus it offers a valuable tool for the benchmarking
of RPA correlation energies.

\begin{table*}
\begin{center}
\caption{\label{tab2}RPA correlation energies of test atoms and molecules
for different basis sets and global extrapolation schemes. 
The last lines report the mean error (ME), the mean absolute error (MAE),
the mean absolute relative error (MARE), the maximum absolute deviation (MAD)
from reference, and the standard deviation of each data set. 
All results are in mHa.}
\begin{ruledtabular}
\begin{tabular}{lrrrrrrrr}
      & \multicolumn{3}{c}{67-extrapolation}& &\multicolumn{3}{c}{56-extrapolation} & \\
\cline{2-4}\cline{6-8}
   & $1/(n+d)^3$ & $1/(n+d)^4$ & $1/n^\alpha$ & & $1/(n+d)^3$ & $1/(n+d)^4$ & $1/n^\alpha$ & Ref.\\
Optimized param. & $d_{opt}=-1.33$ & $d_{opt}=0.37$ & $\alpha_{opt}=3.78$ & & $d_{opt}=-1.17$ & $d_{opt}=0.25$ & $\alpha_{opt}=3.82$ & \\
\hline
H &    -20.9 &    -20.9 &     -20.9 & &   -20.8 &     -20.8 &     -20.8 &  -21.0$\pm$0.1 \\
C &   -292.1 &   -292.1 &    -292.1 &  &  -292.0 &    -292.0 &    -292.0 &  -292.8$\pm$0.7 \\
N &   -337.5 &   -337.5 &    -337.5 &  &  -337.8 &    -337.8 &    -337.8 &  -337.3$\pm$0.6 \\
O &   -431.0 &   -431.0 &    -431.0 &  &  -430.2 &    -430.2 &    -430.2 &  -431.8$\pm$0.9 \\
F &   -523.1 &   -523.1 &    -523.1 &  &  -522.1 &    -522.1 &    -522.1 &  -524.3$\pm$1.3 \\
Ne &   -602.0 &   -602.0 &    -602.0 & &   -601.8 &    -601.8 &    -601.8 &  -602.2$\pm$1.3 \\
H$_2$&  -81.1 &    -81.1 &     -81.1 & &    -81.0 &     -81.0 &     -81.0 & -81.2$\pm$0.2 \\
NH &   -410.4 &   -410.4 &    -410.4 & &   -410.2 &    -410.2 &    -410.2 &  -410.5$\pm$0.6 \\
NH$_2$ &   -480.3 &   -480.3 &    -480.3 & &   -480.0 &    -480.0 &    -480.0 &  -480.6$\pm$0.7 \\
CH$_4$ &   -501.4 &   -501.4 &    -501.4 & &   -501.3 &    -501.3 &    -501.3 &  -501.5$\pm$0.6 \\
NH$_3$ &   -544.9 &   -544.9 &    -544.9 & &   -545.3 &    -545.3 &    -545.3 &  -544.6$\pm$0.7 \\
H$_2$O &   -581.9 &   -581.9 &    -581.9 &  &  -581.8 &    -581.8 &    -581.8 &  -581.9$\pm$0.9 \\
FH &   -602.0 &   -602.0 &    -602.0 &  &  -601.7 &    -601.7 &    -601.7 &  -602.3$\pm$1.1 \\
C$_2$H$_2$ &   -774.5 &   -774.5 &    -774.5 & &   -774.0 &    -774.0 &    -774.0 &  -775.0$\pm$1.1 \\
CN &   -792.8 &   -792.8 &    -792.8 &  &  -793.2 &    -793.2 &    -793.2 &  -792.5$\pm$1.0 \\
HCN &   -819.2 &   -819.1 &    -819.2 &  &  -819.4 &    -819.4 &    -819.4 &  -818.9$\pm$1.0 \\
CO &   -843.4 &   -843.4 &    -843.4 &  &  -843.9 &    -843.8 &    -843.9 &  -843.0$\pm$1.2 \\
N$_2$ &   -855.4 &   -855.4 &    -855.4 &  &  -855.5 &    -855.4 &    -855.5 &  -855.4$\pm$1.2 \\
C$_2$H$_4$ &   -852.1 &   -852.0 &    -852.0 & &   -851.5 &    -851.5 &    -851.5 &  -852.6$\pm$1.1 \\
HCO &   -888.8 &   -888.8 &    -888.8 & &   -889.0 &    -889.0 &    -889.0 &  -888.6$\pm$1.2 \\
H$_2$CO &   -933.1 &   -933.1 &    -933.1 &  &  -933.3 &    -933.3 &    -933.3 &  -933.0$\pm$1.2 \\
O$_2$ &  -1001.9 &  -1001.9 &   -1001.9 & &  -1002.0 &   -1001.9 &   -1001.9 &  -1001.8$\pm$1.5 \\
H$_3$COH &  -1009.0 &  -1009.0 &   -1009.0 & &  -1009.1 &   -1009.0 &   -1009.0 &  -1009.0$\pm$1.3 \\
HOOH &  -1108.0 &  -1107.9 &   -1108.0 &  & -1108.0 &   -1107.9 &   -1108.0 &  -1108.0$\pm$1.6 \\
F$_2$ &  -1163.0 &  -1163.0 &   -1163.0 & &  -1162.4 &   -1162.3 &   -1162.4 &  -1163.6$\pm$2.4 \\
      &          &          &           & &          &           &           & \\
ME & 0.1 & 0.2 & 0.2 & & 0.2 & 0.3 & 0.2 & \\
MAE & 0.3 & 0.3 & 0.3 & & 0.6 & 0.6 & 0.6 & \\
MARE & 0.07\% & 0.07\% & 0.07\% & & 0.13\% & 0.13\% & 0.13\% & \\
MAD & 1.1 & 1.2 & 1.1 & & 2.2 & 2.2 & 2.2 & \\
Std. Dev. & 0.4 & 0.4 & 0.4 & & 0.8 & 0.8 & 0.8 & \\
\end{tabular}
\end{ruledtabular}
\end{center}
\end{table*}

\section{Two-point extrapolations}
In this section we seek for globally optimized parameters $d$ and $\alpha$
to be used in Eq. (\ref{e1}), (\ref{e2}), and (\ref{e4}), capable of
yielding accurate RPA energies with respect to reference values.
The use of global parameters in the extrapolation formulas has in fact
several advantages in practical applications: (1) it allows to compute 
the CBS energy through two-point extrapolation formulas instead of
using a fitting procedure; (2) with a careful optimization of the global 
parameters it allows to effectively include (on average) high-order
effects into the extrapolation, so that smaller basis sets can be
used; (3) it produces more systematic errors, favoring
error cancellation in many applications.
The search for global parameters is also motivated in the present study
by the observation that in the individual fittings of RPA correlation
energies performed in the previous section we observed that for each
formula all the optimal values of $d$ or $\alpha$ were contained
in a rather narrow range. Thus, it is conceivable that ``average''
values of the parameters may exist which produce accurate estimates of the
CBS limit.

For our purpose we consider the two-point extrapolation formulas
\begin{eqnarray}
\label{e5}
E_\infty & = & \frac{E_n(n+d_{opt})^3-E_m(m+d_{opt})^3}{(n+d_{opt})^3-(m+d_{opt})^3} \\
\label{e6}
E_\infty & = & \frac{E_n(n+d_{opt})^4-E_m(m+d_{opt})^4}{(n+d_{opt})^4-(m+d_{opt})^4} \\
\label{e7}
E_\infty & = & \frac{E_nn^{\alpha_{opt}}-E_mm^{\alpha_{opt}}}{n^{\alpha_{opt}}-m^{\alpha_{opt}}}\ , 
\end{eqnarray}
which are derived directly from Eqs. (\ref{e1}), (\ref{e2}), and (\ref{e4}),
respectively, considering two basis sets of cardinal number $n$ and $m$.

The optimization of the global parameters $d_{opt}$ and $\alpha_{opt}$ was
performed using either $n=7$, $m=6$ or $n=6$, $m=5$, and minimizing
the mean absolute error with respect to the reference data of Tab.
\ref{tab1}. 
The use of a smaller basis set ($n=5$, $m=4$) resulted instead into a failure of 
the minimization procedure, because no reasonable compromise value
was found for $d_{opt}$ and $\alpha_{opt}$), reflecting
the difficulty of the V4Zcv basis set to correctly describe the RPA correlation.

The values of the optimized parameters as well the corresponding results for 
the test systems and global statistics are reported in Tab. \ref{tab2}.
We see that all methods yield similar results in good agreement with the
reference values. The differences between the various extrapolation formulas
as well as between data obtained using a 67-extrapolation and a 
56-extrapolation are negligible ($<0.4$ mHa). This indicates
that the optimized global parameters can in fact describe very well the  
correlation convergence behavior, effectively taking into account
higher-order contributions. Concerning general trends, we observe that
56-extrapolations appear to slightly underestimate in general the RPA 
correlation energy, while this effect is reduced for 67-extrapolations.
We stress however, that because the differences between different methods and 
with reference data is well below our estimated accuracy for the benchmark set,
no quantitative conclusions can be drawn from the results of Tab. \ref{tab2},
and to practical purposes all the methods must be considered completely 
equivalent.

As a final note it is instructive to highlight that the global optimization of 
the parameters provides also valuable indications on the convergence behavior
of different formulas. Thus, the values obtained from the optimization
procedure can be used to shed some light on the quality of different
extrapolation schemes. For example, in Refs. 
\cite{bakowies07,bakowies07_2}
it is indicated how $\alpha_{opt}$ provides a measure for 
the rate of convergence of the extrapolation. In our case, the values found
for $\alpha_{opt}$ are greater than the ideal asymptotic value of 3, 
indicating the need for an overweight of the energy obtained
with the largest basis set with respect to the other one in the extrapolation
formula (\ref{e7}),
similarly to that found for CCSD calculations \cite{bakowies07}. 
This means that for the considered basis sets  
the rate of change of the correlation energy with increasing basis set 
is larger than it should be asymptotically, i.e. the basis sets are rather 
inadequate to describe the CBS limit. This is not surprising for
RPA calculations that are well known to converge very difficultly to the
CBS limit and nicely explains the fact that for 56-extrapolation a
slightly larger value of $\alpha_{opt}$ was fond than for 67-extrapolation.

\begin{figure}
\includegraphics[width=0.9\columnwidth]{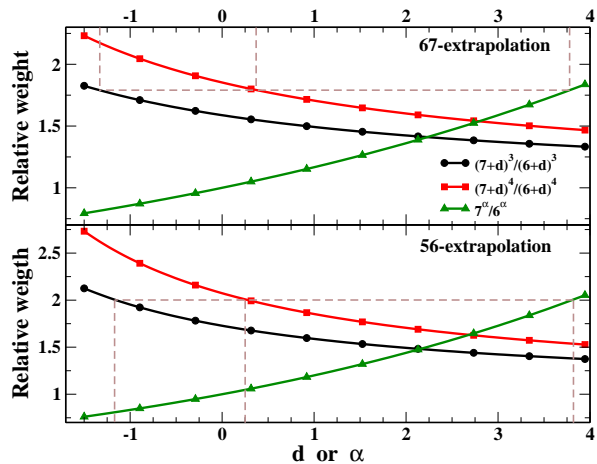}
\caption{\label{fig2} Relative weights (Eq. (\ref{e8})) of the energies used in Eqs. (\ref{e5}), (\ref{e6}), and (\ref{e7}), for different values of the $d$ and $\alpha$ parameters. The dashed lines indicate the location of optimized parameters.}
\end{figure}

For the other two formulas (Eqs. (\ref{e5}) and (\ref{e6})) 
similar considerations apply. 
In fact, it is possible to see that all the optimized parameters provide,
in the extrapolation procedure,
the same relative (over)weight of the energy obtained with the largest basis
set with respect to the other one in all formulas. This is made evident in Fig.
\ref{fig2}, where we plot the ratios 
\begin{equation}\label{e8}
\frac{(n+d)^3}{(m+d)^3} \quad ; \quad \frac{(n+d)^4}{(m+d)^4} \quad ; \quad \frac{n^\alpha}{m^\alpha} 
\end{equation}
for $n=7$, $m=6$ (top panel), $n=6$, $m=5$ (lower panel), at several values
of $d$ or $\alpha$.
The coincidence of the relative weights for different formulas provides thus a 
rationale for the fact that all formulas yield practically the same results
(note that input energies are the same in all formulas at the same level of 
extrapolation).
Inspection of the figure also indicates that the relative weights obtained
from optimized expressions (1.78 and 1.98 for 67-extrapolation and 
56-extrapolation, respectively) are not far from the ``ideal'' ones,
obtained considering $\alpha=3$ (1.60 and 1.74, respectively).

\section{Semiempirical extrapolation from small basis sets}
As discussed in the previous section, for 
smaller basis sets it is not possible 
to find accurate global parameters to extrapolate the RPA correlation energy
to the CBS limit. This problem is clearly related to the inadequacy of the
V4Zcv basis set which makes the progression to
V5Zcv results not regular enough to be extrapolated with a global formula.
The problem can be however partially circumvented by  
employing a suitable technique specificaly developed for
extrapolation from small basis sets.

In wave function theory, e.g. for CCSD(T) calculations, two main strategies
are used to this end: either considering effective empirical parameters for the
extrapolation  \cite{bakowies07,bakowies07_2,truhlar98,fast99} or
considering the relatively inexpensive MP2/CBS limit plus a
$\Delta$CCSD(T) correction evaluated with a smaller basis set \cite{hobza08}.
This latter approach is however not straightforward to extend to RPA calculations,
because it is based on the fact that CCSD(T) and MP2
energies display a very similar basis set dependence. This similarity
is not verified by RPA and MP2 energies, which instead have
a markedly different basis-set behavior (e.g. for the Ne atom passing from
the TZcv to the QZcv basis set the energy changes by 60, 32, and 31 mHa 
for RPA, CCSD(T), and MP2 respectively; changing from the QZcv to the 
5Zcv level the change is 25, 11, and 12 mHa for RPA, CCSD(T), 
and MP2 respectively).

Therefore, in this paper we follow the first strategy and,
following a similar approach as that
proposed by Bakowies \cite{bakowies07,bakowies07_2} for MP2 and 
CCSD calculations, we introduce atom-dependent
parameters which account for the irregular behavior in the progression of 
V4Zcv-V5Zcv energies.
The CBS energy is defined as
\begin{equation}\label{e9}
E_\infty \simeq \frac{E_nn^\gamma - E_mm^\gamma}{n^\gamma - m^\gamma}
\end{equation}
where $n=5$, $m=4$, and
\begin{equation}\label{e10}
\gamma = \frac{\sum_{i=1}^Nn_i\gamma_i}{\sum_{i=1}^Nn_i}
\end{equation}
with the index $i$ running over all atoms in the system, $N$ the total
number of atoms, and $n_i$ the number of electrons of the $i$-th atom.
The atomic parameters $\gamma_i$ are fixed by fitting Eq. (\ref{e9}) 
for single atoms and a minimal set of
homodimers (H$_2$, N$_2$, O$_2$,F$_2$) 
except for carbon where C$_2$H$_2$ was considered. 
Considering only atomic data is in fact not sufficient to yield
accurate and balanced results for all the systems (the mean absolute error 
and the mean relative errors are twice as big as in the case of the minimal
trainig set and the maximum deviation is 4.7 mHa for HOOH), while a large
training set would be likely to lead to artifacts in the fitting. The minimal set 
considered here provides instead a good accuracy limiting as much as possible an arbitrary
selection of reference systems.

The optimized parameters and the resulting RPA
correlation energies are reported in Tab. \ref{tab3}.
\begin{table}
\begin{center}
\caption{\label{tab3} Atomic parameters ($\gamma_i$), extrapolated
RPA correlation energies ($E_c^{RPA}$), differences with
respect to reference energies ($\Delta E^{ref}$), and ratio between 
$\Delta E^{ref}$ and the expected accuracy of each reference energy ($\delta$), 
for the systems of the test set of Tab \ref{tab1}. The last lines report 
the mean error (ME), the mean absolute error (MAE),
the mean absolute relative error (MARE), the maximum absolute deviation (MAD)
from reference, and the standard deviation the extrapolated eenergies.
All results are in mHa.}
\begin{ruledtabular}
\begin{tabular}{lrrrr}
System & $\gamma_i$ & $E_c^{RPA}$ & $\Delta E^{ref}$ & $\Delta E^{ref}/\delta$ \\
\hline
H & 3.10 & -20.9 & 0.1 & 1.3 \\
C & 3.25 & -292.2 & 0.6 & 0.9 \\
N & 3.35 & -337.7 & -0.5 & -0.8 \\
O & 3.23 & -430.9 & 1.0 & 1.1 \\
F & 3.15 & -523.3 & 1.0 & 0.7 \\
Ne & 3.28 & -602.2 & 0.0 & 0.0 \\
H$_2$ &  & -81.2 & 0.0 & -0.1 \\
NH &  & -410.4 & 0.1 & 0.1 \\
NH$_2$ &  & -480.5 & 0.2 & 0.2 \\
CH$_4$ &  & -502 & -0.5 & -0.8 \\
NH$_3$ &  & -547.2 & -2.5 & -3.6 \\
H$_2$O &  & -583.3 & -1.3 & -1.5 \\
FH &  & -603.4 & -1.1 & -1.0 \\
C$_2$H$_2$ &  & -774.9 & 0.1 & 0.1 \\
CN &  & -793.4 & -1.0 & -1.0 \\
HCN &  & -819.8 & -0.9 & -0.9 \\
CO &  & -844.8 & -1.8 & -1.5 \\
N$_2$ &  & -855.2 & 0.2 & 0.1 \\
C$_2$H$_4$ &  & -852.5 & 0.1 & 0.1 \\
HCO &  & -890 & -1.4 & -1.2 \\
H$_2$CO &  & -934.7 & -1.7 & -1.4 \\
O$_2$ &  & -1003.1 & -1.3 & -0.9 \\
H$_3$COH &  & -1010.7 & -1.7 & -1.3 \\
HOOH &  & -1110.1 & -2.1 & -1.3 \\
F$_2$ &  & -1165.1 & -1.5 & -0.6 \\
      &   &          &      & \\
ME    &   &          & -0.7 & \\
MAE   &   &          & 0.9  & \\
MARE  &   &          & 0.16\% & \\
MAD   &   &          & 2.6 & \\
Std. Dev. & &        & 1.0 & \\
\end{tabular}
\end{ruledtabular}
\end{center}
\end{table}

The method performs indeed rather well, yielding a mean absolute error
of only 1 mHa, in line with the estimated accuracy of the reference set,
a mean absolute relative error of 0.16\%. Only in nine cases
over twenty five the extrapolation errors exceed the intrinsic accuracy 
of the reference energies (i.e. $\Delta E^{ref}/\delta> 1$), 
while in all cases the extrapolated energies
improve over the raw data obtained at the V5Zcv level and even with respect
to the raw V7Zcv data.

The present semiempirical extrapolation scheme can thus be a valuable tool for RPA
calculations on large systems, as it allows to achieve a good accuracy at
a relatively small computational cost. We recall in fact that, since 
RPA-correlation calculations scale as $O(N^6)$ 
(which can be reduced to $O(N^4\log N)$ 
using the resolution of the identity (RI) technique \cite{eshuis10}),
where $N$ is the number of basis functions, and a 
cc-pV$n$Z basis set contains approximately
\begin{equation}
N = \frac{(n+1)\left(n+\frac{3}{2}\right)(n+2)}{3}
\end{equation}
basis functions \cite{halkier98}, the computational time required
by the semiempirical approach is about $204^6/91^6\sim$127 (30 when RI
is used) times faster, i.e. two orders of magnitude less, than a
V7Zcv calculation and still $140^6/91^6\sim$13 (6 when RI is used) 
times faster, i.e. one order of magnitude less, than a V6Zcv calculation.

\section{Atomization correlation energies}
To conclude our work we provide in this section a short discussion
on RPA correlation atomization energies, 
that are defined as energy differences between a molecule
and its constituent atoms. 
These can be easily constructed 
from the data of Tabs. \ref{tab1}, \ref{tab2}, and \ref{tab3} or
alternatively, for methods using a two-point extrapolation formula
with global (i.e. system independent) parameters, they can be
obtained by applying the extrapolation formula directly to the
raw atomization energies (i.e computed with two given basis sets). 
The statistics of atomization energies, using as reference data the 
atomization energies obtained from the best estimated RPA energies 
of Tab. \ref{tab1}, are reported in Tab. \ref{tab4}.
\begin{table}
\begin{center}
\caption{\label{tab4} Statistics for the atomization energies computed with different approaches with respect to the best estimated RPA results (see Tab. \ref{tab1}): ME, mean error; MAE, mean absolute error; MARE, mean absolute relative error; MAD, maximum absolute deviation; Std. Dev., standard deviation. All results are in mHa. For two-point extrapolation schemes the results are reported only once, since all the procedures are essentially superimposable for atomization energies.}
\begin{ruledtabular}
\begin{tabular}{lrrrrr}
Method & ME & MAE & MARE & MAD & Std. Dev. \\
\hline
V4Zcv & -1.5 & 1.7 & 1.27\% & 5.0 & 1.6 \\
V5Zcv &  0.2 & 1.2 & 0.93\% & 3.0 & 1.4 \\
V6Zcv &  0.9 & 1.3 & 1.05\% & 2.6 & 1.4 \\
V7Zcv &  0.9 & 1.2 & 0.95\% & 2.3 & 1.1 \\
56-extr. & 1.5 & 1.8 & 1.38\% & 3.4 & 1.5 \\
67-extr. & 1.0 & 1.1 & 0.84\% & 2.0 & 0.8 \\
semiemp. & 1.0 & 1.0 & 0.75\% & 2.2 & 0.6 \\
\end{tabular}
\end{ruledtabular}
\end{center}
\end{table}

The inspection of Tab. \ref{tab4} reveals that several approaches
are capable to yield very good results, with mean absolute errors
close to 1 mHa and mean relative errors smaller than 1\%.
In particular, all the calculations using raw basis-set results (except V4Zcv)
perform remarkably well, showing a significantly better performance
with respect to the case of absolute energies, thanks to the cancellation
of systematic errors in the present case.
Interestingly little or no benefit comes instead from the use of 
extrapolation techniques. In fact, 67-extrapolation results are in 
line with V7Zcv results, while 56-extrapolation atomization energies
are even worst that V5Zcv ones and of similar quality as the
V4Zcv results. This result shall be rationalized in terms of
an inevitable small but random computational noise that affects
the extrapolation data (especially if small basis sets are used as a base
for the extrapolation). The small inaccuracies of extrapolated data
can be in fact almost irrelevant for
absolute energies (below 1 mHa), but can easily sum to several mHa
in the case of atomization energies (especially for many atoms molecules).
Finally, we remark the good performance of the semiempirical
extrapolation method, which yields atomization energies in good agreement with
the reference ones, with a mean absolute relative error of only 0.75\%
and very small maximum absolute deviation and standard deviation.

The results of Tab. \ref{tab4} show that 
the quality of each extrapolation procedure to yield
accurate atomization energies cannot be directly
inferred from the results obtained for absolute energies, because
atomization energies are energy differences and thus a complex
error propagation can occur. As we showed this effect is more
important for low-level approaches and is somehow amplified for extrapolated results,
because of the presence of unsystematic errors due to the numerical procedure.
It is thus finally important to note that this issue applies 
also for our reference atomization energies, 
that were obtained from the best estimated RPA absolute energies of Tab. \ref{tab1},
although in this case we may expect the effect to be rather limited due to the
high quality of the extrapolation.
Nevertheless, we have to remark that possibly errors of few mHa are plausible for 
some systems, so that in general the set cannot be fully considered
as an accurate benchmark set of atomization energies, although it could be
safely considered as reference for a qualitative discussion
of the trends obtained from different extrapolation methods.

\section{Conclusions}
The slow basis set convergence of the RPA correlation energy is a major
problem for the development and the application of the method in
quantum chemistry and solid-state physics.
In fact, a careful assessment of the RPA methodology and related methods
(e.g., RPA+ \cite{ruzsinszky10}, RPA+SE \cite{ren11}), requires
the existence of a set of benchmark absolute and relative RPA-correlation
energies as well as the availability of accurate extrapolation techniques.
Practical applications, on the other hand, would greatly
benefit from efficient extrapolation schemes with well calibrated levels of
confidence.

Surprisingly however only little effort has been dedicated in literature to 
study the CBS limit of RPA correlation energies and assess the effectiveness of
different extrapolation procedures in this context.
In this paper we aimed at pushing this work one step further and
considered the problem in a more systematic way, constructing a 
benchmark set of reference RPA-correlation energies and studying 
the behavior of different extrapolation schemes in the RPA framework.
Our work can be summarized in the following conclusions:
\begin{itemize}
\item The extremely slow convergence of the RPA correlation 
energy with the basis set dimension prevents 
the computation of highly accurate reference energies.
Already at the V7Zcv level in fact the energies can be incorrect up to 10 mHa. 
Nevertheless, we showed that a careful extrapolation procedure allows to
construct 
an accurate benchmark set with a well defined estimation
of the uncertainty. In the present work we estimate our reference data to 
have an accuracy close to 1 mHa for most of the systems.
\item We analyzed several two-point extrapolation formulas against our 
benchmark set of absolute RPA correlation energies. It turns out that  
good results can only be achieved if global parameters are optimized in Eqs.
(\ref{e5}), (\ref{e6}), and (\ref{e7}), as to effectively reproduce
high-order terms in the theoretical asymptotic energy 
expansion ($\propto 1/n^3$).
When this procedure is followed all the considered formulas provide equivalent
results. Moreover,
with a careful optimization of the global parameters very good results
can also be achieved from 56-extrapolation procedures.
\item RPA correlation energies evaluated at the V4Zcv level appear to 
be of too low quality to serve as a basis in an 
extrapolation procedure with globally optimized parameters. 
This fact indicates that in the case of 45-extrapolations 
high-order terms beyond the $1/n^3$ one, play
a crucial role and are thus dominant in the asymptotic expansion. 
Therefore, RPA energies obtained at the quadruple-zeta level 
of theory cannot be considered
well converged with respect to the one-particle expansion 
in basis set and shall not be used in applications.
Nevertheless, the introduction of atomic-based parameters 
within a semiempirical extrapolation scheme can strongly 
reduce the problem and provide extrapolated RPA absolute energies
of good quality (within 2 mHa from the reference).
The semiempirical extrapolation method can thus be 
considered as a good tool for the CBS extrapolation in
RPA applications on large systems. 
\item When energy differences are considered, an unpredictable 
error propagation occurs and accurate CBS results are even 
more hard to achieve than the absolute correlation energies. 
However, because an error cancellation occurs 
for the systematic errors (e.g. systematic underestimation of 
the energy by incomplete basis sets) we finally
found that most approaches yield results that agree within 2 mHa.
Because no highly accurate reference data exist to assess such 
small differences, 
no clear preference can be expressed for any of the considered 
approaches and 2 mHa ($\sim1.2$ kcal/mol, $\sim0.05$ meV) 
must be considered the level of confidence of state-of-the-art RPA 
calculations on atomization energies. Therefore,
caution must be always employed when small energy difference are 
considered.
\end{itemize}

We mention finally that, following the encouraging results of the
present study, further investigations will be needed to assess
in detail the basis set-dependence of RPA energies and the
accuracy of extrapolation schemes for energy differences
(e.g. atomization energies), which
are a main quantity in many practical computational studies.
In fact, when energy differences are considered, an unpredictable 
error propagation may occur and accurate CBS results are  
harder to achieve than the absolute correlation energies,
despite the former can be directly derived as an algebraic sum of
the latter. 
From our calculations it can be easily seen indeed that all the approaches
considered in this work yield 
very similar atomization energies for the molecules of the
test set, with differences mostly below 2-3 mHa. However, 
no clear trend can be identified between the different approaches,
so that no highly accurate reference data can be established to assess such 
small differences.
This traces back to the fact that systematic errors 
(e.g. systematic underestimation of the energy by incomplete basis sets) 
may be expected to cancel out in this case, but extrapolation procedures 
may introduce in general unsystematic errors which worsen the results with
respect to the raw data obtained from the corresponding basis sets.

\begin{acknowledgements}
We thank TURBOMOLE GmbH for providing us with the TURBOMOLE
program package, and M. Margarito for technical support. This work was
funded by the ERC Starting Grant FP7 Project DEDOM, Grant Agreement No. 207441.
\end{acknowledgements}

\end{document}